\documentclass[hyper]{JHEP} 

\usepackage{epsfig}

\newcommand\fverb{\setbox\pippobox=\hbox\bgroup\verb}

\newcommand\fverbdo{\egroup\medskip\noindent%

            \fbox{\unhbox\pippobox}\ }

\newcommand\fverbit{\egroup\item[\fbox{\unhbox\pippobox}]}

\newbox\pippobox

\title{Note About Classical Dynamics of
Pure Spinor String on $AdS_5\times S_5$
Background }

\author{J. Kluso\v{n}
 \footnote{On leave from Masaryk University, Brno}\\
Dipartimento di Fisica,\\
Universita' di Roma \& Sezione di Roma 2, Tor Vergata \\
Via della, Ricerca, Scientifica, 1 00133  Roma   ITALY\\
E-mail:
\email{Josef.Kluson@roma2.infn.it}}
 \preprint{
\hepth{0603228}}
\abstract{We will discuss some properties
of the pure spinor string on the $AdS_5\times S_5$ background. 
Using the classical Hamiltonian
analysis we will show  that
the  vertex operator for the massless state that is
in the cohomology of the BRST charges
describes on-shell fluctuations 
around $AdS_5\times S_5$ background.}

\keywords{string theory}

\def\tr{\mathrm{Tr}}

\def\str{\mathrm{Str}}

\def\pb  #1{\left\{#1\right\}}

\newcommand{\bP}{{\bf P}}
\newcommand{\hp}{\hat{p}}

\newcommand{\mH}{\mathcal{H}}

\newcommand{\htheta}{\hat{\theta}}

\newcommand{\hd}{\hat{d}}
\newcommand{\halpha}{\hat{\alpha}}
\newcommand{\hbeta}{\hat{\beta}}
\newcommand{\hdelta}{\hat{\delta}}
\newcommand{\hgamma}{\hat{\gamma}}
\newcommand{\hlambda}{\hat{\lambda}}
\newcommand{\hkappa}{\hat{\kappa}}
\newcommand{\hw}{\hat{w}}
\newcommand{\hN}{\hat{N}}

\newcommand{\hmu}{\hat{\mu}}

\newcommand{\com}[1]{\left[#1\right]}
\newcommand{\oz}{\overline{z}}

\newcommand{\tP}{\tilde{P}}
\newcommand{\hP}{\hat{P}}
\newcommand{\hpi}{\hat{\pi}}

\newcommand{\uc}{\underline{c}}
\newcommand{\ud}{\underline{d}}
\newcommand{\uf}{\underline{f}}
\newcommand{\ue}{\underline{e}}

\newcommand{\hLambda}{\hat{\Lambda}}
\newcommand{\hGamma}{\hat{\Gamma}}
\begin{document}
\section{Introduction and summary}
The string action that posses
manifest space time supersymmetry, is famous Green-Scwarz
(GS) action \cite{Green:1983wt}. 
However, no one has succeeded in making
a covariant quantisation of GS action. The source
of the difficulty is impossibility to achieve
the separation of the fermionic first class
and second class constraints associated with
local $\kappa$ symmetry in a manifestly covariant
way. As in ten dimensions the smallest covariant
spinor corresponding to Majorana-Weyl spinor has
16 real components, $8$ first class and $8$ second
class constraints (that arise in heterotic
or Type I GS superstings) do not fit into such
covariant spinor representation separately.

It was soon realised that second
class constraints are present in the GS action
due to the definition of the conjugate momenta
to $\theta$. By using proposal of Faddeev and
Frandkin one could turn second class constraints
into the first call ones by adding further
fields but upon quantisation one now obtained
infinite set of ghost-for-ghosts that in the
end is infinite \cite{Kallosh:1989yv,Gates:1989hg}. 

Few years ago Berkovits has proposed an 
interesting approach to covariant quantisation
of superstrings, using pure spinors
\cite{Berkovits:2000fe,Berkovits:2000ph,Berkovits:2000nn,
Berkovits:2001us}. 
The Berkovits approach seems to be the
right directions for covariant quantisation
of the GS action as was shown on many examples
in the past. In fact, it appears that the
statement that the pure spinor approach 
provides a super-Poincare covariant quantization of 
superstring theories is correct but deserves
a warning. Since this approach is based on
very simple form of the BRST operator 
is is not obvious how it can be obtained
by gauge fixing a reparametrisation invariant 
worldsheet action. On the other hand there
exist some proposals to find more geometrical
version of the pure spinor formalism
\cite{Grassi:2004we,Grassi:2004nz,Guttenberg:2004ht,
Grassi:2003kq,Gaona:2005yw,Oda:2005sd,
Aisaka:2005vn,Chesterman:2004xt,Aisaka:2003mw,Matone:2002ft,Oda:2001zm}
and for recent discussion of these approaches
we recommend the paper 
\cite{Berkovits:2005bt} where new 
promising formulation of the pure spinor
formalism was also proposed. 
In summary, the super-Poincare covariant
formalism can be applied in many situations,
for example in the computation of the
multi loop scattering amplitudes
\cite{Berkovits:2004px,Anguelova:2004pg,Berkovits:2005ng,Mafra:2005jh}
and the formulation of the quantum
superstring in the $AdS_5\times S_5$
Ramond-Ramond background.

The superstring worldsheet action 
in $AdS_5\times S_5$ Ramond-Ramond 
background can be studied at the classical
level using either the GS
formalism \cite{Metsaev:1998it}
 or the pure spinor formalism
\cite{Berkovits:2000yr}. Since
$AdS_5\times S_5$ is solution
of type IIB supergravity the worldsheet
action is classically
$\kappa$-invariant in the
GS formalism and is classically BRST
invariant in the pure spinor 
formalism \cite{Berkovits:2001ue}.
Moreover, it was even shown in 
\cite{Berkovits:2004xu} that
the BRST invariance of the
pure spinor action persists at
the quantum level. 

However even if it was shown
that the pure spinor action in
$AdS_5\times S_5$ with Ramond-Ramond
flux leads to well
defined CFT this theory has not
been solved yet. The main difficulty
 is 
in the fact that we cannot separate
worldsheet fields to their
holomorphic and antiholomorphic
parts. The
peculiar property of the
Ramond-Ramond background
can be nicely seen in the
analysis of string
theory on $AdS_3\times S_3$ background.
While the case with 
 nonzero $NS$ two form field
 is well known understood
 the  the analysis of the
$AdS_3\times S_3$ with Ramond-Ramond
flux is much more involved 
\cite{Berkovits:1999im}.

Due to these facts  we mean
that it is useful to collect as
far as informations about the
pure spinor string theory in $AdS_5\times
S_5$ background as possible. 
In particular  we mean that 
the classical canonical analysis
of the covariant string in 
$AdS_5\times S_5$ could be helpful
for the study of the quantum
conformal field theory. This
paper is the first step in the
study of the  properties of the
classical covariant string in the 
$AdS_5\times S_5$ background.

The rest of this  paper is organised as follows.
In the next section (\ref{second})
we review the pure spinor approach
in the flat background with emphasis
on the worldsheet covariant 
formulation that allows classical
Hamiltonian analysis. 

In section (\ref{third}) we 
will study the properties of
the pure spinor action in general
background that was introduced
in \cite{Berkovits:2001ue}. We
again formulate this action
in manifest covariant formalism
even if we have to stress that
we work on the flat worldsheet.
We express the BRST charge
in terms of the canonical
variables of the theory and
then we determine the action
of this charge on the 
function on the extended
phase space and that can
be interpreted as the vertex
operator for massless 
state of the theory. Then
we show that the requirement
that this vertex 
is in the cohomology of
the BRST charges leads to
the result that this function
describes massless fluctuations
around given background that
obey appropriate linearised
equations of motion. 

In section (\ref{fourth})
we begin to discuss the
classical pure spinor action
on $AdS_5\times S_5$. We
review the constructions
of BRST charges  performed 
in \cite{Berkovits:2000yr,Berkovits:2002zv}
with emphasis on the   manifest
worldsheet covariant formulation. We also
determine the equations of
motions for matter and
ghost fields. Then we
express the BRST charge
in terms of canonical
variables and we will 
calculate its action on
function defined on the phase
space and that is classical
analogue of the vertex operator in
the standard treatment.
  We again
argue that the requirement
that this function belongs
to the cohomology of BRST
charges implies that
these vertex operators 
correspond to on-shell
fluctuations on the
$AdS_5\times S_5$ background.

The summary of this paper is
as follows. We explicitly determine
the action of the BRST charge
in general background on
function on the phase space and we will
argue that in general the
requirement that this function
is in the cohomology of the
BRST charges implies that
this function describes on-shell
fluctuations around given
supergravity background. 
We obtain the result that
is in the agreement with
the action of the  BRST charge
on these vertex operators that
 was suggested in 
\cite{Berkovits:2000yr}. 

We would like also stress that
this paper gives only modest
contribution to the 
 study of the classical
canonical formalism of pure
spinor action in $AdS_5\times S_5$.
As a next step we would
like to   determine
the algebra of the currents
 that defines given action.
Then  the knowledge of this algebra
will be helpful for construction
of more general functions on the
phase spaces that  could serve
as an analogues of  the vertex operators
for massive states on the $AdS_5\times
S_5$ background. This work is
currently progress. 

\section{Pure Spinor Superstring in
Flat Background}\label{second}
In this section we  give a brief
review of the pure spinor formalism in
flat spacetime. For more 
details, see
\cite{Berkovits:2002zk,Grassi:2005av,
Grassi:2003cm,Grassi:2002sr,Grassi:2002xf}.

In a flat background, the worldsheet
action  in the pure spinor formalism
takes the form 
\footnote{We work in units $2\pi\alpha'=1$.}
\begin{eqnarray}\label{Sclas}
S=-\int
d^2x\sqrt{-\eta}(\frac{1}{2}\eta^{\mu\nu}
\partial_\mu Y^m\partial_\nu Y^n
\eta_{mn}+p_{\mu\alpha}
P^{\mu\nu}\partial_\nu \theta^\alpha
+\hp_{\mu\halpha}\tP^{\mu\nu}
\partial_\nu\htheta^{\halpha}+
\nonumber \\
+w_{\mu\alpha}P^{\mu\nu}
\partial_\nu\lambda^\alpha
+\hw_{\mu\alpha}\tP^{\mu\nu}
\partial_\nu \hlambda^{\halpha}) \ , 
\nonumber \\
\end{eqnarray}
where 
\begin{eqnarray}
\eta_{\mu\nu}=\mathrm{diag}(-1,1) \ , 
\sqrt{-\eta}=\sqrt{-\det \eta} \ , 
\nonumber \\
P^{\mu\nu}=
\eta^{\mu\nu}-\epsilon^{\mu\nu} \ ,
\tP^{\mu\nu}=
\eta^{\mu\nu}+\epsilon^{\mu\nu} \ ,
P^{\mu\nu}=
\tP^{\nu\mu}  \ , 
\nonumber \\
\epsilon^{\mu\nu}=
\frac{e^{\mu\nu}}{\sqrt{-\eta}} \ ,
e^{01}=-e^{10}=1 \ . \nonumber \\
\end{eqnarray}
We label the  worldsheet
 with variables
$x^\mu \ , x^0=t \ ,
x^1\equiv x$. The flat spacetime indices are labeled
with  $ m,n=0,\dots,9$, spinor indices
are labeled with 
$\alpha,\halpha=1,\dots,16$. Finally
$\lambda^\alpha, \hlambda^{\halpha}$
are bosonic ghosts satisfying the
pure spinor conditions
\begin{equation}
\lambda^\alpha(\gamma^m)_{\alpha\beta}
\lambda^\beta=0 \ , 
\hlambda^{\halpha}(\gamma^m)_{\halpha
\hbeta}\hlambda^{\hbeta}=0 \ , 
\mathrm{for \ m= \ 0 \ to \ 9 } \ , 
\end{equation}
where $\gamma^m_{\alpha\beta} \ , 
\gamma^m_{\halpha\hbeta}$ are $16\times 16$ symmetric
matrices that are off-diagonal components
of the $32\times 32$ gamma matrices.
Finally, as we will see below,
$w_{\mu\alpha} \ , \hw_{\mu\halpha}$
are related to the momentum conjugate
to $\lambda^\alpha, \hlambda^{\halpha}$. 
At the same time $p_{\mu\alpha} \ , 
\hp_{\mu\halpha}$ are related to the momenta
conjugate to $\theta^{\alpha} \ , 
\htheta^{\halpha}$. 

Fundamental object of the pure
spinor formalism are BRST charges
$Q_R,Q_L$ 
\begin{equation}\label{qlrc}
Q_L=\int dx j^0_L \ , 
Q_R=\int dx j^0_R \ , 
\end{equation}
where
\begin{eqnarray}\label{jlr}
j^\mu_L=\lambda^\alpha
\tP^{\mu\nu}d_{\nu\alpha}=\lambda^\alpha
d_{\nu\alpha}P^{\nu\mu} \ , \nonumber \\
j^\mu_R=\hlambda^{\halpha}
P^{\mu\nu}\hd_{\nu\halpha}=
\hlambda^{\halpha}\hd_{\nu\halpha}
\tP^{\nu\mu}  \  . \nonumber \\
\end{eqnarray}
Since   $Q_{R,L}$ have to be
time independent 
 the currents
(\ref{jlr}) are conserved 
\begin{equation}\label{conseq}
\partial_\mu j^\mu_{R,L}=0 \ .
\end{equation}
Even if the notation used
above is slightly unconventional
 it is equivalent
to the standard one.  Namely, 
performing the Wick rotation on
the worldsheet and then 
introducing  chiral 
variables $z,\oz$ it is easy
to see that the only nonzero
components of the projectors $P,\tP$
are $P^{z\oz}=1 \ , \tP^{\oz z}=1$.
Then the  equation
(\ref{conseq}) is equal to
\begin{equation}
\partial_{\oz} (\lambda^\alpha 
d_{\alpha z})=0 \ , 
\partial_{z} (\hlambda^{\halpha} 
\hd_{\halpha \oz})=0 \ 
\end{equation}
that allow us to construct
two conserved charges
\begin{equation}
  Q_L=\frac{1}{2\pi i}
\oint dz \lambda^\alpha
d_{\alpha z} \ ,
Q_R=\frac{1}{2\pi i}
\oint d\oz \hlambda^{\halpha}
\hd_{\halpha\oz} \ 
\end{equation}
that are the standard
forms of the BRST charges
used in the literature. Since
in what follows we will study
the pure spinor action using
Hamiltonian formalism we
use the convention given
in (\ref{qlrc}). 
We must also stress that
in
the flat spacetime
the objects  $d_{\mu\alpha},
\hd_{\mu\halpha}$   take the form
\begin{eqnarray}
d_{\mu\alpha}=
p_{\mu\alpha}
+(i\partial_\mu Y^m
+\frac{1}{2}\theta^\alpha
(\gamma^m)_{\alpha\beta}
\partial_\nu\theta^\beta+
\frac{1}{2}\htheta^{\halpha}
(\gamma^m)_{\halpha\hbeta}
\partial_\nu\htheta^{\hbeta})
(\gamma_m\theta)_\alpha \ ,
\nonumber \\
\hd_{\mu\halpha}=
\hp_{\mu\halpha}
+(i\partial_\mu Y^m
+\frac{1}{2}\theta^\alpha
(\gamma^m)_{\alpha\beta}
\partial_\nu\theta^\beta+
\frac{1}{2}\htheta^{\halpha}
(\gamma^m)_{\halpha\hbeta}
\partial_\nu\htheta^{\hbeta})
(\gamma_m\htheta)_{\halpha} \ .
\nonumber \\
\end{eqnarray}
The physical states in the pure
spinor formalism are defined
as vertex operators of ghost number
$(1,1)$ in the cohomology of the
nilpotent BRST charges.
The massless states are constructed
from zero modes only and hence
are represented by the vertex
operator
\begin{equation}\label{Uflat}
U=\lambda^{\alpha}\hlambda^{\hbeta}
A_{\alpha\hbeta}(Y^m,\theta,\htheta) \ ,
\end{equation}
where $A_{\alpha\hbeta}$ is bispinor
$N=2$ $D=10$ superfield which
depends on on the worlsheed zero
modes of $y^m,\theta^\alpha$ and
$\htheta^{\halpha}$ only.

In terms of the BRST analysis
of the classical theory this
condition translates  to the requirement that 
the gradded Poisson bracket
\footnote{Definition of the
gradded Poisson bracket will
be given in next section.}
 between $Q_{L,R}$ 
and $U$ is equal to zero
\begin{equation}\label{QLRU}
\pb{Q_L,U}=\pb{Q_R,U}=0 \ . 
\end{equation}
It is clear that shift of the 
function $U$ in the form
\begin{equation}
\delta U=\pb{Q_L,\Lambda}+\pb{Q_R,\hLambda} \ ,
\end{equation}
where $\pb{Q_R,\Lambda}=\pb{Q_L,\hLambda}=0$ 
does not change the equations 
(\ref{QLRU}).  
Applying these conditions to $U$ given in
(\ref{Uflat}) and if we also define
$\Lambda,\hLambda$ as 
\begin{equation}
\Lambda=\hlambda^{\halpha}\hat{\Gamma}_{\halpha} \ , 
\hLambda=\lambda^{\alpha}\Gamma_{\alpha}
\end{equation}
and using the fact that pure spinors satisfy
\begin{equation}\label{purespinorcon}
\lambda^\alpha\lambda^\beta=
\frac{1}{3840}(\lambda \gamma^{mnpqr}
\lambda)\gamma^{\alpha\beta}_{mnpqr} \ , 
\hlambda^{\halpha}\hlambda^{\hbeta}=
\frac{1}{3840}(\hlambda \gamma^{mnpqr}
\hlambda)\gamma^{\halpha\hbeta}_{mnpqr} \ 
\end{equation} 
we find that $A_{\alpha\hbeta}$ satisfy
the conditions
\begin{equation}\label{eqf}
\gamma^{\alpha\beta}_{mnpqr}D_{\alpha}
A_{\beta\hgamma}=\gamma^{\halpha\hbeta}_{mnpqr}
D_{\halpha}A_{\gamma\hbeta}=0 \ 
\end{equation}
that together with gauge transformations
\begin{equation}\label{gf}
\delta A_{\beta\hgamma}=
D_{\beta}\hat{\Gamma}_{\hgamma}+
D_{\hgamma}\Gamma_{\beta}  \ , 
\gamma^{\alpha\beta}_{mnpqr}
D_{\alpha}\Gamma_{\beta}=\gamma^{\halpha\hbeta}_{mnpqr}
D_{\halpha}\hat{\Gamma}_{\hgamma}=0 \ ,
\end{equation}
where  $D_\alpha=\frac{\partial}{\partial \theta^\alpha}+
(\gamma^m\theta)_\alpha \partial_m \ , 
D_{\halpha}=\frac{\partial}{\partial \htheta^{\halpha}}
+(\gamma\htheta)_{\halpha}\partial_m$ are 
the supersymmetric derivatives of flat $N=2$
D=10 superspace.
Performing now the analysis
as in  
\cite{Berkovits:2000yr}
one can argue that (\ref{eqf}) and (\ref{gf}) 
correctly reproduce the 
Type IIB supergravity
spectrum in flat spacetime. 
 
After this brief review of the
main properties of the pure spinor
formalism in flat spacetime we
focus in the next section on the
formulation of the pure spinor
string in general background.

\section{Pure spinor action in 
General Background}\label{third}
The general form of the
pure spinor action in Type IIA, IIB theories
was given in 
\cite{Berkovits:2001ue,Berkovits:2002zk}.
In what follows we concentrate
on Type IIB theory and we  write
this pure spinor action in
the manifest worldsheet covariant 
notation using   $\eta^{\mu\nu} \ ,
\epsilon^{\mu\nu}$ and projectors 
$P^{\mu\nu} , \tP^{\mu\nu}$. 
Then the pure
spinor action in general background
takes the form 
\footnote{We omit the Fradkin-Tseytlin
term $\int \Phi(Z)r$ where $\Phi$
is dilaton superfield and $r$ is
worldsheet curvature.}
\begin{eqnarray}\label{genactcov}
S=-\int d^2x\sqrt{-\eta}
\left[\frac{1}{2}
\eta^{\mu\nu}g_{\mu\nu}
-\frac{\epsilon^{\mu\nu}}{2}
b_{\mu\nu}
+\right. \nonumber \\
\left.+
P^{\alpha\hbeta}d_{\alpha\mu}
 P^{\mu\nu}
\hd_{\hbeta\nu}+d_{\alpha\mu}
P^{\mu\nu} \partial_\nu
Z^ME^\alpha_M(Z)
 +\hd_{\halpha\mu}
\tP^{\mu\nu}\partial_\nu Z^M
E^{\halpha}_M(Z) +\right. \nonumber \\
\left.+ \lambda^{\alpha}w_{\beta\mu}
P^{\mu\nu}\partial_\nu 
Z^M\Omega_{M\alpha}^{ \
\beta}+
\hat{\lambda}^{\halpha}\hat{w}_{\hbeta\mu}
\tP^{\mu\nu}
\partial_\nu Z^M
\hat{\Omega}_{M\halpha}^{\hbeta}
+ \right. \nonumber \\
\left. + C_{\alpha}^{\beta \hat{\gamma}}
\lambda^\alpha w_{\beta\mu}\tP^{\mu\nu}
\hd_{\hgamma\nu}+\hat{C}^{\hbeta\gamma}_{\halpha}
\hat{\lambda}^{\halpha}\hw_{\hbeta}
P^{\mu\nu}d_{\gamma\nu}
+S^{\beta \hdelta}_{\alpha
\hgamma}\lambda^\alpha w_{\beta\mu}
\hlambda^{\hgamma}P^{\mu\nu}
\hw_{\hdelta\nu}\right]+S_\lambda+ S_{\hlambda}
\ ,
\nonumber \\
S_\lambda=-\int d^2x\sqrt{-\eta}
w_{\mu\alpha}P^{\mu\nu}\partial_\nu
\lambda^\alpha \ , 
S_{\hlambda}=
-\int d^2x\sqrt{-\eta}
\hw_{\mu\halpha}\tP^{\mu\nu}\partial_\nu
\hlambda^{\halpha} \ , 
\nonumber \\
\end{eqnarray}
where we have defined
\begin{equation}
g_{\mu\nu}=\partial_\mu Z^ME_M^a
\partial_\nu Z^NE_N^b\eta_{ab} \ , 
b_{\mu\nu}=
\partial_\mu Z^M\partial_\nu Z^NB_{MN} \ ,
\end{equation}
and where $M=(m,\mu \ , \hmu)$ are curved
superspace indices, 
$Z^M=(x^m,\theta^\mu \ , \theta^{\hmu})$,
$A=(a,\alpha,\halpha)$ are tangent superspace
indices, $S_\lambda \ , S_{\hlambda}$
are the flat actions for the pure spinor
variables. Finally
\begin{equation}
E_M^\alpha \ , E_M^{\halpha},
\Omega_{M\alpha}^{ \ \beta} \ ,
\hat{\Omega}_{M\halpha}^{\ \hbeta} \ ,
P^{\alpha\hbeta} \ ,
C_{\alpha}^{\beta\hgamma} \ ,
\hat{C}^{\hbeta\gamma}_{\halpha} \ ,
S^{\beta \hdelta}_{\alpha\hgamma} \ 
\end{equation}
are background spacetime fields.
Note also  that $d_{\mu\alpha} \ ,
\hd_{\mu\halpha}$ should be treated
as an independent variables since
$p_{\mu\alpha} \ , \hat{p}_{\mu\halpha}$
do not appear explicitly.

As in the flat space the fundamental
object of  the pure spinor formalism in
the general background are the BRST operators
$Q_L,Q_R$ that again take the form
\begin{equation}\label{QLRg}
Q_L=\int dx\lambda^\alpha 
d_{\mu \alpha}P^{\mu 0} \ , 
Q_R=\int dx\hlambda^{\halpha}
\hd_{\mu\halpha}\tP^{\mu 0} \ .
\end{equation}

Our goal is to show that
even at the classical level
the requirement that the function 
that depends on $Z^M$ only is
in the cohomology of the BRST 
charges (\ref{QLRg}) 
implies  that this function
describes on-shell massless fluctuations
around given supergravity background.
Our analysis can be considered as
 modest contribution to the
more general study performed in 
\cite{Berkovits:2001ue}.

To use the classical formalism
we have to  express $d_{\mu\alpha} \ , 
\hd_{\mu\halpha}$ in terms of canonical
variables of the extended phase
space with coordinates $(Z^M,P_M,\lambda^\alpha,
\pi_{\alpha} \ , \hlambda^{\halpha} \ ,
\hpi_{\halpha})$, where
  $P_M$ is  a momentum conjugate to
$Z^M$ and $\pi_{\alpha}\ , \hpi_{\halpha}$
are momenta conjugate to $\lambda^\alpha$ and
$\hlambda^{\halpha}$ respectively.
Then using the  graded 
Poisson brackets we will calculate the action
of this BRST operators (\ref{QLRg}) on
the function on the extended phase
space. 

To proceed we firstly determine
 the momentum
conjugate to $Z^M$.
The conjugate momentum $P_M$ 
is defined as the left  variation of
the action with respect to $
\partial_0 Z^M$ 
\begin{eqnarray}\label{PMg}
P_M=\frac{\delta^L S}{\delta
\partial_0Z^M}=
E_M^a\eta_{ab}E^b_N\partial_0Z^N
+\partial_1Z^N B_{MN}
-\nonumber \\
+ E_M^\alpha d_{\alpha\mu}
P^{\mu 0}+
 E^{\halpha}_M d_{\halpha \mu}
\tP^{\mu 0}-
\nonumber \\
- \lambda^{\alpha}w_{\alpha\mu}
P^{\mu 0}\Omega_{M\alpha}^{ \
\beta}-
\hat{\lambda}^{\halpha}\hat{w}_{\hbeta\mu}
\tP^{\mu 0}
\hat{\Omega}_{M\halpha}^{\hbeta} \ .
 \nonumber \\
\end{eqnarray}
Then we define canonical
graded Poisson bracket between
$P_M$ and $Z^M$ in the form
\begin{equation}\label{cpb}
\pb{Z^M(x),P_N(y)}=
(-1)^{|M|}\delta^M_N\delta(x-y) \ , 
\end{equation}
where   $|X|$ is
equal to one if $X$ is Grassmann
odd and $|X|=0$ if $X$ is Grassman
even.
The gradded Poisson
bracket is defined as follows.
Consider physical system
with canonical
bosonic variables $q^i$ with
corresponding 
conjugate momenta $p_i$ together
with fermionic variables
$\omega^\alpha$ with
conjugate momenta $\pi_\alpha$
\footnote{Since we work on the
extended phase space that contains
ghosts and their conjugate momenta
we implicitly presume that they are
included in the set of bosonic $q^i$
or  fermionic $\omega^\alpha$ 
variables given above. For example, since
pure spinor ghosts and their conjugate
momenta are bosonic they are included
in the set of  variables $q^i$.}.
Then the graded Poisson
brackets involving fermionic
systems is defined as
\begin{eqnarray}\label{gpb}
\pb{F,G}=
\left[\frac{\partial F}{\partial g^i}
\frac{\partial G}{\partial \pi_i}
-\frac{\partial F}{\partial \pi_i}
\frac{\partial G}{\partial g^i}\right]
+(-1)^{|F|}
\left[\frac{\partial^LF}{\partial
\omega^\alpha}\frac{\partial^LG}{
\partial \pi_\alpha}+
\frac{\partial^LF}{\partial \pi_\alpha}
\frac{\partial^LG}{\partial \omega^\alpha}
\right] \ , \nonumber \\
\end{eqnarray}
where the superscript $L$ 
represents left derivation
\footnote{The relation between
left and right derivative
can be found as follows. Let
$F$ is a function of Grassmann
parity $|F|$ defined on 
superspace labeled with $Z^M$.  Since
$dF(Z)=dZ^M\partial^L_MF=\partial^R_MFdZ^M$
we obtain that 
left and right derivatives of $F$
are related as $
(-1)^{|M||M+F|}\partial^L_M F
=\partial^R_M F$.}. In what
follows we will consider the derivative
from the left only and for that
reason we omit the superscript $L$ on
the sign of the partial derivative. 

 Now if  we
introduce the notation  $Z^M=(q^i,\omega^\alpha)$ and
$P_M=(p_i,\pi_\alpha)$ then
(\ref{gpb}) can be written
in a  more symmetric form 
\begin{equation}\label{gpbs}
\pb{F,G}=
(-1)^{|F||M|}
\left[\frac{\partial^L F}{\partial Z^M}
\frac{\partial^L G}{\partial P_M}
-(-1)^{|M|}\frac{\partial^L F}{\partial P_M}
\frac{\partial^L G}
{\partial Z^M}\right] \ . 
\end{equation}
Note that these graded 
Poisson brackets 
 obey the relation
\begin{eqnarray}\label{pbfg}
\pb{F,G}=
-(-1)^{|F||G|}\pb{G,F} \ 
\nonumber \\
\end{eqnarray}
and
\begin{eqnarray}\label{pbfgh}
\pb{F,GH}=
\pb{F,G}H+(-1)^{|H||G|}\pb{F,H}G \ . 
\nonumber \\
\end{eqnarray}
Finally we introduce  the momenta 
conjugate to $\lambda^\alpha \ ,
\hlambda^{\halpha}$ as
\begin{equation}
\pi_{\mu\alpha}=\frac{\delta S}{\delta
\partial_0\lambda^\alpha}=
-w_{\mu\alpha}P^{\mu 0} \ ,
\hpi_{\mu\halpha}=
\frac{\delta S}{\delta \partial_0
\hlambda^{\halpha}}=
-\hw_{\mu\halpha}\tP^{\mu 0}
\end{equation}
with the canonical commutation
relations
\begin{equation}\label{ghostpb}
\pb{\lambda^\alpha(x),\pi_{\beta}(y)}
=\delta^\alpha_\beta\delta(x-y) \ ,
\pb{\hlambda^{\halpha}(x),\hpi_{\hbeta}(y)}
=\delta^\alpha_\beta\delta(x-y) \ .
\end{equation}
Note that $\pi_{\alpha} \ , \hpi_{\halpha}$
carry the ghost number $(-1,0),(0,-1)$ respectively.

If we now return to (\ref{PMg})
and multiply  this expression
with $E^\alpha_M$ from the left  and  use the
fact that $E_A^ME_M^B=\delta_A^B$ 
we obtain
\begin{eqnarray}\label{dPpi}
d_{\alpha \mu}P^{\mu 0}=
E_\alpha^M
\left(-P_M+\partial_1Z^N B_{MN}
-\Omega_{M\gamma}^{ \
\beta} \lambda^{\gamma}w_{\beta\mu}
P^{\mu 0}-
\hat{\Omega}_{M\halpha}^{\hbeta}
\hat{\lambda}^{\halpha}\hat{w}_{\hbeta\mu}
\tP^{\mu 0}\right)=
\nonumber \\ 
E_\alpha^M
\left(-P_M+\partial_1Z^N
B_{MN}
+\Omega_{M\gamma}^{ \
\beta} \lambda^{\gamma}\pi_\beta+
\hat{\Omega}_{M\halpha}^{\hbeta}
\hat{\lambda}^{\halpha}\hpi_{\hbeta}
\right) \ .
\nonumber \\
\end{eqnarray}
Then  using (\ref{ghostpb})
we easily get
\begin{eqnarray}\label{dpla}
\pb{d_{\alpha\mu}P^{\mu 0}(x),
\lambda^\beta(y)}=
-
E_\alpha^M
\Omega_{M\gamma}^\beta\lambda^\gamma(y)
\delta(x-y) \ , \nonumber \\
\pb{d_{\alpha\mu}P^{\mu 0}(x),
\hlambda^{\hbeta}(y)}=
-E_\alpha^M
\hat{\Omega}_{M\hgamma}^{\hbeta}
\hlambda^{\hgamma}(y)
\delta(x-y) \ . \nonumber \\
\end{eqnarray}
These relations will be important
in the next subsection.

On the other hand if we multiply
(\ref{PMg}) with $E_{\halpha}^M$
from the left and proceed in
the same way as above we get
\begin{equation}
\hd_{\halpha\mu}
\tP^{\mu 0}=E_{\halpha}^M
\left(-P_M+
\partial_1 Z^NB_{MN}
+\Omega_{M\gamma}^{ \
\beta} \lambda^{\gamma}\pi_\beta+
\hat{\Omega}_{M\halpha}^{\hbeta}
\hat{\lambda}^{\halpha}\hpi_{\hbeta}
\right) \ 
\nonumber \\ 
\end{equation}
and also
\begin{eqnarray}
\pb{\hd_{\alpha\mu}\tP^{\mu 0}(x),
\lambda^\beta(y)}=
-E_{\halpha}^M
\Omega_{M\gamma}^\beta\lambda^\gamma(y)
\delta(x-y) \ , \nonumber \\
\pb{\hd_{\halpha\mu}\tP^{\mu 0}(x),
\hlambda^{\hbeta}(y)}=
-E_{\halpha}^M
\hat{\Omega}_{M\hgamma}^{\hbeta}
\hlambda^{\hgamma}(y)
\delta(x-y) \ . \nonumber \\
\end{eqnarray}
\subsection{Equation of motion
for massless fields in general background}
Now we will argue 
that the ghost number
$(1,1)$  function on 
the extended phase space that
 depends on 
$Z^M$ only and that   is in the  cohomology of
the BRST charges
$Q_L,Q_R$ describes
on-shell states of the  massless fluctuations
around given general background.

At the first step we define
the function-vertex  of 
the ghost number    $(1,1)$ in the form
\begin{equation}\label{V11}
V^{(1,1)}=\lambda^\alpha
\hlambda^{\halpha} A_{\alpha\halpha}(Z)\ .
\end{equation}
Then we demand that this function
is in the cohomology of the
BRST operators (\ref{QLRg}), namely
\begin{equation}\label{QLRgV}
\pb{Q_L,V^{(1,1)}}=
\pb{Q_R,V^{(1,1)}}=0 \  
\end{equation}
and also that the operator (\ref{V11}) satisfies
the gauge invariance
\begin{equation}\label{dVg}
\delta V^{(1,1)}=
\pb{Q_L,\Lambda^{(0,1)}}+
\pb{Q_R,\hLambda^{(1,0)}} \ , 
\end{equation}
where
\begin{equation}\label{eqgp}
\pb{Q_R,\Lambda^{(0,1)}}=
\pb{Q_L,\hLambda^{(1,0)}}=0 \ ,
\end{equation}
and where $\Lambda^{(0,1)}\ , 
\hLambda^{(1,0)}$ are  superfields
of the ghost numbers $(0,1),(1,0)$ respectively
so that can be written as 
\begin{equation}\label{gpex}
\Lambda^{(0,1)}=\hlambda^{\halpha}
\hGamma_{\halpha}(Z) \ , 
\hLambda^{(1,0)}=
\lambda^\alpha \Gamma_{\alpha}(Z) \ .
\end{equation} 
In order to calculate the Poisson
bracket between $Q_L$ and $V$ 
we use  (\ref{cpb}) that implies 
\begin{equation}
\pb{P_M(x),V_{\alpha\halpha}(Z(y))}=
-(-1)^{|M|}\partial_M V_{\alpha
\halpha}(Z(x))
\delta(x-y) \ . 
\end{equation}
If we also use (\ref{dpla})
we easily obtain
\begin{eqnarray}
\pb{\lambda^\gamma d_{\gamma\mu}
(x)P^{\mu 0},\lambda^\alpha\hlambda^{\halpha}
A_{\alpha\halpha}(y)}=
\nonumber \\
\lambda^\gamma\lambda^\alpha\hlambda^{\halpha}
E_\gamma^M
\left((-1)^{|M|}\partial_MA_{\alpha\halpha}-
\Omega_{\alpha M}^\beta A_{\beta\halpha}-
\hat{\Omega}_{\halpha M}^{\hbeta}
A_{\alpha\hbeta}
\right)\delta(x-y) \ . 
\nonumber \\
\end{eqnarray}
Using this result
and also (\ref{purespinorcon})
we get that (\ref{QLRgV})
is equivalent to the
following equation
\begin{equation}\label{eqgL}
\gamma_{mnpqr}^{\gamma\alpha}
E_\gamma^M
\left((-1)^{|M|}\partial_MA_{\alpha\halpha}-
\Omega_{\alpha M}^\beta A_{\beta\halpha}-
\hat{\Omega}_{\halpha M}^{\hbeta}
A_{\alpha\hbeta} 
\right)=0 \ . 
\end{equation}
In the same way we  proceed
with $Q_R$ and we get 
\begin{equation}\label{eqgR}
\gamma_{mnpqr}^{\hgamma\halpha}
E_{\hgamma}^M
\left((-1)^{|M|}\partial_MA_{\alpha\halpha}-
\Omega_{\alpha M}^\beta A_{\beta\halpha}-
\hat{\Omega}_{\halpha M}^{\hbeta}
A_{\alpha\hbeta}
\right)=0 \ . 
\end{equation}
We can argue, following  
\cite{Berkovits:2000yr} that
these equations correctly
describe the on-shell fluctuations
around given supergravity  solutions.

Let us now consider the 
action of the BRST current
on the operator of the ghost 
number $(M,N)$ that has the form
\begin{equation}\label{Phimn}
\Phi=\lambda^{\alpha_1}\dots
\lambda^{\alpha_M}\hlambda^{\hbeta_1}
\dots \hlambda^{\hbeta_N}
A_{\alpha_1\dots\alpha_M
\hbeta_1\dots \hbeta_N}
(Z) \ . 
\end{equation}
Since the  function  (\ref{Phimn})
does not depend
on $P_M$ it is again easy to calculate
the action of the BRST charges
(\ref{QLRg}) on it and we get
\begin{eqnarray}\label{QPhi}
\pb{Q_L,\Phi(y)}=
\lambda^{\kappa}\lambda^{\alpha_1}
\dots \lambda^{\alpha_M}
\hlambda^{\hbeta_1}
\dots \hlambda^{\hbeta_M}
\nabla_{\kappa}A_{\alpha_1
\dots\alpha_M\hbeta_1\dots\hbeta_N}(y) 
\equiv \tilde{\Phi}^{M+1,N}(y) 
\ , \nonumber \\
\pb{Q_R,\Phi(y)}=
\lambda^{\kappa}\lambda^{\alpha_1}
\dots \lambda^{\alpha_M}
\hlambda^{\hat{\kappa}}\hlambda^{\hbeta_1}
\dots \hlambda^{\hbeta_M}
\nabla_{\hat{\kappa}}A_{\alpha_1
\dots\alpha_M\hbeta_1\dots\hbeta_N}(y)
\equiv
\tilde{\Phi}^{M,N+1}(y) 
\ , \nonumber \\
\end{eqnarray}
where $\tilde{\Phi}^{M+1,N}(Z)$
is a function of the ghost number
$M+1,N$   while   $\Phi^{M,N+1}(Z)$ 
is the function of the ghost number
$M,N+1$.
In the previous expression  the action
of $\nabla_\kappa$ on $A_{\alpha_1\dots\alpha_M
\hbeta_1\dots \hbeta_N}$ has the form
\begin{eqnarray}
\nabla_{\kappa}A_{\alpha_1
\dots\alpha_M\hbeta_1\dots\hbeta_N}=
E_\kappa^M
((-1)^{|M|}\partial_M A_{\alpha_1
\dots\alpha_M\hbeta_1\dots\hbeta_N}-\nonumber \\
\Omega_{M\alpha_1}^{\gamma}A_{\gamma\alpha_2\dots
\alpha_M\hbeta_1\dots\hbeta_N}\dots-
 \Omega_{M\alpha_M}^\gamma A_{\alpha_1\dots
\alpha_{M-1}\gamma\hbeta_1\dots\hbeta_N}
-
\nonumber \\
\Omega_{M\hbeta_1}^{\hgamma}
A_{\alpha_1\dots\alpha_M\hgamma\hbeta_2\dots\hbeta_N}
\dots -
\Omega_{M\hbeta_N}^{\hgamma}
A_{\alpha_1\dots\alpha_N\hbeta_1\dots\hbeta_{N-1}\hgamma}) \ . 
\nonumber \\
\end{eqnarray}
Using the expression  (\ref{QPhi}) we
easily obtain that the gauge
transformation of $V^{(1,1)}$ 
(\ref{dVg}) implies
\begin{equation}
\delta A_{\alpha\halpha}=
\nabla_{\alpha}\hGamma_{\halpha}+
\nabla_{\halpha}\Gamma_{\alpha} \ , 
\end{equation}
where we have also used (\ref{gpex}). Note
also that (\ref{eqgp}) and (\ref{QPhi})
together with the pure spinor constraint
(\ref{purespinorcon}) imply
\begin{equation}
\gamma^{\alpha\beta}_{mnpqr}
\nabla_{\alpha}\Gamma_{\beta}=
\gamma^{\halpha\hbeta}_{mnpqr}
\nabla_{\halpha}\hGamma_{\hbeta}=0 \ .
\end{equation}
All these relations are correct form of the
equations of motion and gauge symmetries
of the massless fluctuations around given
background. 

However it is important to stress that
the pure spinor formalism is consistent
in case when the BRST operators $Q_L,Q_R$ are
nilpotent:
\begin{equation}
\pb{Q_L,Q_L}=\pb{Q_R,Q_R}=0 \ 
\end{equation}
and also when $Q_L,Q_R$ have vanishing
Poisson bracket among themselves
\begin{equation}
\pb{Q_L,Q_R}=0 \ .
\end{equation}
At the same time we have to demand that
the currents $j^{\mu}_{L,R}$ are
separately conserved.
As was shown in 
\cite{Berkovits:2001ue} all these
conditions imply  
a set of constraints on the background
superfields. We we will not review
these calculations and recommend
the paper \cite{Berkovits:2001ue} for
more details.

\section{Pure spinor action in
$AdS_5\times S_5$}\label{fourth}
In this section we review
the construction of the pure
spinor action in $AdS_5\times S_5$
with emphasis on the covariant
worldsheet formulation. 

As in the Matsaev-Tseytlin
GS action in and $AdS_5\times
S_5$ background  
\cite{Metsaev:1998it} the action
in the pure spinor formalism 
\cite{Berkovits:2000fe,Berkovits:2000yr,Berkovits:2004xu}
is constructed form the left-invariant
currents 
\begin{equation}\label{JA}
J^A_\mu=(g^{-1}\partial_\mu g)^A=
\partial_\mu Z^ME^A_M \  ,
\end{equation}
where  $g$ takes values in $PSU(2,2|4)/SO(4,1)\times
SO(5)$,
$A=(c,[cd],c',[c'd'],\alpha,\halpha)$
range over tangent space indices of
the super-Lie algebra of $PSU(2,2|4)$
where $(c,[cd])$ describe the
$SO(4,2)$ isometries of $AdS_5$ with
$c=0,\dots,4$ and $(c',[c'd'])$ 
describe the $SO(6)$ isometries of
$S^5$ with $c'=5\dots,9$. We also
preserve the notation  $\alpha$ 
and $\halpha$ for the two sixteen -
 component Majorana-Weyl spinors. 
Finally, $Z^M$ label the target
superspace and hence $M$ ranges
over the curved superspace 
indices $(m=0,\dots,9 \ , 
\mu=1,\dots,16 \ , \hmu=1,\dots,16$). 

It was also argued in
\cite{Berkovits:2000fe}, following
\cite{Berkovits:1999zq} 
 that the only nonzero components of $B_{\mu\nu}=
E^A_\mu E^B_\nu B_{AB}$ and $F^{\alpha\hbeta}$ are
\begin{equation}\label{Balphabeta}
B_{\alpha\hbeta}=
-(-1)^{|\alpha||\hbeta|}B_{\hbeta\alpha}=
B_{\hbeta\alpha}=\frac{1}{2}(ng_s)^{1/4}\delta_{\alpha
\hbeta} \ , 
F^{\alpha\hbeta}=\frac{1}{(ng_s)^{1/4}}
\delta^{\alpha\hbeta} \ ,
\end{equation}
where $n$ is the value of the Ramond-Ramond flux,
$g_s$ is the string coupling constant and
$\delta_{\alpha\hbeta}=
(\gamma^{01234})_{\alpha\hbeta}$. 

Now we are ready to write the
pure spinor action in $AdS_5\times S_5$.
Firstly, using  (\ref{JA})
we can write
the embedding of $g_{\mu\nu}$ 
 to the worldsheet of
the string as 
\begin{equation}\label{gemb}
g_{\mu\nu}=
\partial_\mu Z^ME_M^{\uc}\partial_\nu Z^N
E_N^{\ud}\eta_{\uc\ud}=J^{\uc}_\mu J^{\ud}_\nu 
\eta_{\uc\ud} \ ,  
\end{equation}
where $\uc$ signifies 
either $c$ or $c'$. In the same
way  $[\uc\ud]$
signifies either $[cd]$ or
$[c'd']$. 

In the same way the embedding of
$b_{\mu\nu}$ can be written as
\begin{eqnarray}\label{bemb}
\epsilon^{\mu\nu}B_{\mu\nu}=
\epsilon^{\mu\nu}E_\mu^AE_\nu^BB_{AB}=
(ng_s)^{1/4}J_\mu^\alpha J_\nu^{\hbeta}
\delta_{\alpha\hbeta} \nonumber \\
\end{eqnarray}
using (\ref{Balphabeta}), antisymmetry
of $\epsilon^{\mu\nu}$ and 
the fact that $J^\alpha,J^{\hbeta}$ are
Grassmann odd.
Then the matter part of the pure spinor action in
$AdS_5\times S_5$ takes the form  
\begin{eqnarray}\label{Smatter}
S=-\int d^2x
\sqrt{-\eta}\left(\frac{1}{2}
\eta^{\mu\nu}g_{\mu\nu}-\frac{\epsilon^{\mu\nu}}{2}
b_{\mu\nu}+F^{\alpha\hbeta}
d_{\alpha\mu}P^{\mu\nu}
\hd_{\hbeta\nu}+\right.
\nonumber \\
\left. E_\nu^\alpha d_{\beta\mu}P^{\mu\nu}+
E_\nu^{\halpha}\hd_{\hbeta\mu}
\tP^{\mu\nu}\right)=
\nonumber \\ 
=-\int d^2x\sqrt{-\eta}
\left(\frac{1}{2}\eta^{\mu\nu}
J_\mu^{\uc} J_\nu^{\ud}\eta_{\uc\ud}
-(ng_s)^{1/4}\frac{\epsilon^{\mu\nu}}{2}
J_\mu^\alpha J_\nu^{\hbeta}
\delta_{\alpha\hbeta}+\right.
\nonumber \\
\left.+\frac{1}{(ng_s)^{1/4}}\delta^{\alpha\hbeta}
d_{\mu\alpha}P^{\mu\nu}
\hd_{\nu\hbeta}+J_\nu^\alpha d_{\alpha\mu}P^{\mu\nu}+
J_\nu^{\halpha}\hd_{\halpha\mu}
\tP^{\mu\nu}\right) \  
\end{eqnarray}
while   the ghost
action  takes the form
\cite{Berkovits:2000fe,Berkovits:2000yr}
\begin{eqnarray}\label{Sghost}
S_{ghost}=
-\int d^2x\sqrt{-\eta}
(w_{\mu\alpha}P^{\mu\nu}
\partial_\nu \lambda^\alpha
+\hw_{\mu\halpha}\tP^{\mu\nu}
\partial_\nu\hlambda^{\halpha}
+\nonumber \\
+N_{\uc\ud\mu}
P^{\mu\nu}J^{[\uc\ud]}_\nu+
\hN_{\uc\ud\mu}
\tP^{\mu\nu}J^{[\uc\ud]}_\nu-
N_{cd\mu}P^{\mu\nu}
\hN^{cd}_\nu+N_{c'd'\mu}P^{\mu\nu}
N^{c'd'}_\nu) \ . 
\nonumber \\
\end{eqnarray}

Since $d_{\alpha\mu} \ , \hd_{\halpha\mu}$ appear
in (\ref{Smatter})  as  auxiliary fields 
they can be integrated out 
and we obtain
\begin{eqnarray}
P^{\mu\nu}d_{\hbeta\nu}=(ng_s)^{1/4}
\delta_{\hbeta\alpha}P^{\mu\nu}J^\alpha_\nu \ ,
\nonumber \\
\tP^{\nu\mu}d_{\mu\alpha}=-(ng_s)^{1/4}\delta_{\alpha
\halpha}\tP^{\nu\mu}J_{\mu}^{\halpha} \ . 
\nonumber \\
\end{eqnarray}
Inserting these expressions back to the
(\ref{Smatter}) we obtain the contribution 
\begin{equation}
-(ng_s)^{1/4}J^{\halpha}_\mu P^{\mu\nu}
\delta_{\halpha\alpha}J_\nu^\alpha
\end{equation}
and consequently the actions
(\ref{Smatter}) and (\ref{Sghost}) together
 take the form 
\begin{eqnarray}
S=-\int d^2x\sqrt{-\eta}
(\frac{1}{2}\eta^{\mu\nu}J_\mu^{\uc} J_\nu^{\uc}
\eta_{\uc\ud}
+(ng_s)^{1/4}\eta^{\mu\nu}
J^\alpha_\mu J^{\hbeta}_\nu 
\delta_{\alpha\hbeta}
+(ng_s)^{1/4}\frac{1}{2}
\epsilon^{\mu\nu}J_\mu^{\alpha} J_\nu^{\hbeta}
\delta_{\alpha\hbeta}
\nonumber \\
-\int d^2x\sqrt{-\eta}
(w_{\mu\alpha}P^{\mu\nu}
\partial_\nu \lambda^\alpha
+\hw_{\mu\halpha}\tP^{\mu\nu}
\partial_\nu\hlambda^{\halpha}
+\nonumber \\
+N_{\uc\ud\mu}
P^{\mu\nu}J^{[\uc\ud]}_\nu+
\hN_{\uc\ud\mu}
\tP^{\mu\nu}J^{[\uc\ud]}_\nu-
(N_{cd\mu}P^{\mu\nu}
\hN^{cd}_\nu-N_{c'd'\mu}P^{\mu\nu}
N^{c'd'}_\nu)) \ . 
\nonumber \\
\end{eqnarray}
If we now perform rescaling
\begin{equation}
J^{\uc}\rightarrow (ng_s)^{1/4}J^{\uc}
\ , J^\alpha\rightarrow  (ng_s)^{1/8}J^{\alpha} \ ,
J^{\halpha}\rightarrow (ng_s)^{1/8}J^{\halpha} \ , 
J^{[\uc\ud]}\rightarrow (ng_s)^{1/4}J^{[\uc\ud]}
\end{equation}
and also 
\begin{equation}
(\lambda,w,\hlambda,\hw)
\rightarrow (Ng_s)^{1/8}(\lambda,w,
\hlambda,\hw) \ , (N,\hN)\rightarrow
(ng_s)^{1/4}(N,\hN) \ 
\end{equation}
we obtain the action in the form
\begin{eqnarray}\label{SADS}
S=-\sqrt{ng_s}\int d^2x\sqrt{-\eta}
\left(\frac{1}{2}\eta^{\mu\nu}J_\mu^{\uc} 
J_\nu^{\ud}\eta_{\uc\ud}
+\eta^{\mu\nu}
J^\alpha_\mu J^{\hbeta}_\nu 
\delta_{\alpha\hbeta}
+\frac{\epsilon^{\mu\nu}}{2}J_\mu^{\alpha} J_\nu^{\hbeta}
\delta_{\alpha\hbeta}+\right.
\nonumber \\
\left.+w_{\mu\alpha}P^{\mu\nu}
\partial_\nu \lambda^\alpha
+\hw_{\mu\halpha}\tP^{\mu\nu}
\partial_\nu\hlambda^{\halpha}
+\right.\nonumber \\
\left.+N_{\uc\ud\mu}
P^{\mu\nu}J^{[\uc\ud]}_\nu+
\hN_{\uc\ud\mu}
\tP^{\mu\nu}J^{[\uc\ud]}_\nu
-N_{cd\mu}P^{\mu\nu}
\hN^{cd}_\nu+N_{c'd'\mu}P^{\mu\nu}
\hN^{c'd'}_\nu   \right) \ .
\nonumber \\
\end{eqnarray}
In what follows we omit
the factor $\sqrt{n g_s }$.

Note that  the action (\ref{SADS}) can be
written also in the form 
\begin{eqnarray}\label{Sha}
S=-\int d^2x\sqrt{-\eta}
(\frac{1}{2}\eta^{\mu\nu}J_\mu^i J_\nu^jG_{ij}
+\frac{\epsilon^{\mu\nu}}{2}J_\mu^{i} J_\nu^{j}
B_{ij}
\nonumber \\
+w_{\mu\alpha}P^{\mu\nu}
\partial_\nu \lambda^\alpha
+\hw_{\mu\halpha}\tP^{\mu\nu}
\partial_\nu\hlambda^{\halpha}
+\nonumber \\
+N_{\uc\ud\mu}
P^{\mu\nu}J^{[\uc\ud]}_\nu+
\hN_{\uc\ud\mu}
\tP^{\mu\nu}J^{[\uc\ud]}_\nu-
N_{cd\mu}P^{\mu\nu}
\hN^{cd}_\nu+N_{c'd'\mu}P^{\mu\nu}
\hN^{c'd'}_\nu) \ ,   
\nonumber \\
\end{eqnarray}
 where 
$i=\uc,\alpha,\halpha$ ad  we have
  defined 
\begin{equation}\label{Gij}
G_{ij}=\left(\begin{array}{ccc}
\eta_{\uc\ud}  & 0 & 0 \\
0 & 0 & \delta_{\alpha\hbeta} \\
0 &-\delta_{\halpha\beta}  & 0 \\
\end{array}\right) \ ,
B_{ij}=\frac{1}{2}\left(\begin{array}{ccc}
0 & 0 & 0 \\
0 & 0 & \delta_{\alpha\hbeta} \\
0 & \delta_{\halpha\beta} & 0 \\
\end{array}\right) \  
\end{equation}
that obey 
\begin{equation}
G_{ij}=(-1)^{|i||j|}G_{ji} \ ,
B_{ij}=-(-1)^{|i||j|}B_{ji} \ .  
\end{equation}
For letter purposes it will be
useful to know the inverse matrix $G^{ij}$
\begin{equation}\label{GijI}
G^{ij}=\left(\begin{array}{ccc}
\eta^{\uc\ud} & 0 & 0 \\
0 & 0 & -\delta^{\alpha\hbeta} \\
0 & \delta^{\halpha\beta}  & 0 \\
\end{array}\right) \ , 
\end{equation}
where $\eta_{\mu\nu}\eta^{\nu\kappa}=
\delta_\mu^\kappa \ ,
\delta_{\alpha\hbeta}\delta^{\hbeta\gamma}=\delta_\alpha^\beta \ ,
\delta^{\halpha\beta}\delta_{\beta\hgamma}=
\delta^{\halpha}_{\hgamma}$. 

The form of the action (\ref{Sha})
will be useful for Hamiltonian
analysis of the pure spinor action. 
\subsection{Alternative form of the
action }
It turns out that for the
study of the equations of motion
of pure spinor string it is
useful to use an alternative
form of the action that was
introduced recently in the paper
\cite{Berkovits:2004xu}.
In the covariant worldsheet formalism
this action takes the form
\begin{eqnarray}\label{Minaction}
S=-\int
d^2x\sqrt{-\eta}\str\left(\frac{1}{2}
\eta^{\mu\nu}\left(J_\mu^{(2)}
J_{\nu}^{(2)}+J_\mu^{(1)}J_\nu^{(3)}
+J_\mu^{(3)}J_\nu^{(1)}\right)+
\right.\nonumber\\
\left.+\frac{\epsilon^{\mu\nu}}{4}
\left(J^{(1)}_\mu J^{(3)}_\nu-
J^{(3)}_\mu J^{(1)}_\nu\right)
\right)
\nonumber \\
-\int d^2x\sqrt{-\eta}
\str(w_{\mu}P^{\mu\nu}
\partial_\nu \lambda
+\hw_{\mu}\tP^{\mu\nu}
\partial_\nu\hlambda
+\nonumber \\
N_{\mu}
P^{\mu\nu}J^{(0)}_\nu+
\hN_{\mu}
\tP^{\mu\nu}J^{(0)}_\nu-
\frac{1}{2}N_{\mu}P^{\mu\nu}\hN_\nu
-\frac{1}{2}\hN_{\mu}\tP^{\mu\nu}
N_\nu)  \ ,
\nonumber \\
\end{eqnarray}
where 
\begin{eqnarray}
J^{(0)}_\mu=(g^{-1}\partial_\mu g)^{[\uc\ud]}T_{[\uc\ud]} \ ,
J^{(1)}_\mu=(g^{-1}\partial_\mu g)^{\alpha}T_{\alpha} \ ,
\nonumber \\
J^{(2)}_\mu=(g^{-1}\partial_\mu g)^{\uc}T_{\uc} \ ,
J^{(3)}_\mu=(g^{-1}\partial_\mu g)^{\halpha}
T_{\halpha} \ , \nonumber \\
w_\mu=w_{\alpha\mu}G^{\alpha\hbeta} T_{\hbeta}
 \ ,
\lambda=\lambda^\alpha T_\alpha  \ ,
\nonumber \\
N_\mu=-\pb{w_\mu,\lambda}=-w_{\mu\beta}G^{\beta\halpha}\lambda^\alpha
\pb{T_{\halpha},T_{\alpha}}=
\nonumber \\
\frac{1}{2}w_{\mu\alpha}(\gamma^{cd})^{\alpha}_\beta 
\lambda^\beta T_{[cd]}
-\frac{1}{2}w_{\mu\alpha}(\gamma^{c'd'})^{\alpha}_\beta 
\lambda^\beta T_{[c'd']}=
\nonumber \\
N_{\mu}^{cd}T_{[cd]}-N_\mu^{c'd'}T_{[c'd']}
 \ ,
\nonumber \\
\hw_\mu=\hw_{\halpha\mu}G^{\halpha\beta}
 T_{\beta}
 \ ,
\hlambda=\hlambda^{\halpha} T_{\halpha}  \ ,
\nonumber \\
\hat{N}_\mu=-\pb{\hw_\mu,\hlambda}= 
\nonumber \\
\hN_{\mu}^{cd}T_{[cd]}-\hN_\mu^{c'd'}T_{[c'd']}
 \ \nonumber \\
\end{eqnarray}
using
\begin{equation}
\pb{T_\alpha,T_{\hbeta}}=
\frac{1}{2}(\gamma^{cd})_{\alpha}^{\gamma}
\delta_{\gamma\hbeta}T_{[cd]}-
\frac{1}{2}(\gamma^{c'd'})_{\alpha}^{\gamma}
\delta_{\gamma\hbeta}T_{[c'd']} \ .
\end{equation}
It can be easily
shown   that 
(\ref{Minaction}) is equivalent to the
action (\ref{SADS})
with the help of the relations 
\begin{equation}
\str(T_{[\uc\ud]}T_{[\ue\uf]})=k_{[\uc\ud][\ue\uf]} \ , 
\str(T_{\uc}T_{\ud})=\eta_{\uc\ud} \ ,
\str (T_\alpha, T_{\hbeta})=G_{\alpha\hbeta} \ , 
\str(T_{\hbeta}T_\alpha)=G_{\hbeta\alpha} \ ,
\end{equation}
where $k_{[ab][cd]}$ is defined
as 
\begin{equation}
k_{[cd][ed]}=\eta_{c[e}\eta_{f]d} \ , 
k_{[c'd'][e'd']}=-\delta_{c'[e'}\delta_{f']d'} 
\ , 
k_{[cd][e'f']}=k_{[c'd'][ef]}=0 \ . 
\end{equation}
At this place we give a short 
review of some properties of $psu(2,2|4)$
algebra.  We represent an element
of this superalgebra by an even supermatrix
of the form
\begin{equation}\label{Gsup}
G=\left(\begin{array}{cc}
A & X \\
Y & B \\ \end{array}\right) \ , 
\end{equation}
where $A$ and $B$ are matrices
with Grassmann even functions while
$X$ and $Y$ are those with Grassmann odd functions,
each representing a $4\times 4$ matrix. 
(An odd supermatrix has the same form 
with $A$ and $B$ consisting of Grassmann
odd functions while $X$ and $Y$ consisting
of Grassmann even functions. An example
of odd supermatrix is $\lambda=
\lambda^\alpha T_\alpha$ that has Grassmann
even functions on off-diagonal blocks). 

An element $G$ of the superalgebra 
$psu(2,2|4)$ is given by a $8\times 8$
matrix (\ref{Gsup}) satisfying 
\begin{equation}
\Sigma A^{\dag}+A\Sigma=0 \ , 
B^{\dag}+B=0 \ , 
X-i\Sigma Y^{\dag}=0 \ , 
\end{equation} 
where $\Sigma=\sigma_3\otimes I_2$ with
$\sigma_3$ standard Pauli matrix and with
$I_2$ representing the identity matrix in $2$
dimensions. 

The essential feature of the superalgebra
$psu(2,2|4)$ is that it admits
a $\mathbf{Z}_4$ automorphism 
such that the condition $\mathbf{Z}_4(H)=H$
determines the maximal subgroup
to be $SO(4,1)\times SO(5)$ that
leads to the definition of the
coset for the sigma model. 
The $\mathbf{Z}_4$ authomorphism
$\Omega$ takes an element of $psu(2,2|4)$
to another $G\rightarrow \Omega(G)$
such that
\begin{equation}
\Omega(G)=\left(\begin{array}{cc}
JA^TJ & -JY^TJ  \\
JX^TJ & JB^TJ \\ \end{array}
\right)  \ , 
J=\left(\begin{array}{cc}
0 & -1 \\
1 & 0 \\ \end{array}\right) \ . 
\end{equation}
Since $\Omega^4=1$ the eigenvalues of
$\Omega$ are $i^p \ , p=0,1,2,3$. Therefore
we can decompose the superalgebra
$G$ as
\begin{equation}
G=\mH_0\oplus\mH_1\oplus\mH_2\oplus \mH_3 \ , 
\end{equation}
where $\mH_p$ denotes the eigenspace of 
$\Omega$  such that if $H\in \mH_p$ then
\begin{equation}\label{Omegaav}
\Omega(H_p)=i^pH_p \ .
\end{equation}
The authomorphism $\Omega$
also implies an important relation
\begin{equation}\label{comhpq}
\com{H_p,H_q}\in \mH_{p+q} \ (\mathrm{mod \ 4}) \ .
\end{equation}
As we have argued above $\Omega(\mH_0)=
\mH_0$ determines $\mH_0=
SO(4,1)\times SO(5)$ and $\mH_0$  is spanned
by generators $T_{[\uc\ud]}$. $\mH_2$ represents
the remaining bosonic elements of the
algebra and it is spanned by  generators $T_{\uc}$.
$\mH_1$ contains
fermionic elements of the algebra 
and it is spanned with generator $T_\alpha$ while
$\mH_3$ contains fermionic elements of
the algebra  and it is spanned with
the generators  
$T_{\halpha}$. 
Then we can write the current $J_\mu$
as
\begin{eqnarray}
J_\mu=J^A T_A=J^{(0)}_\mu+J^{(1)}_\mu+
J^{(2)}_\mu+J^{(3)}_\mu \ , 
\nonumber \\
J^{(0)}_\mu=J^{[\uc\ud]}T_{[\uc\ud]} \ , 
J^{(1)}_\mu=J^{\alpha}_\mu T_\alpha \ , 
\nonumber  \\
J^{(2)}_\mu=J^{\uc}_\mu T_{\uc} \ , 
J^{(3)}_\mu=J^{\halpha}_\mu T_{\halpha} \ ,
\nonumber \\
\end{eqnarray}
where $A=([\uc\ud],\alpha,\uc,\halpha)$
and where $J^\alpha_\mu \ , J^{\halpha}_\mu$
are Grassmann odd functions while 
$J^{[\uc\ud]}_\mu \ , J^{\uc}_\mu$ are
Grassmann even functions. The generators 
$T_A$ satisfy the graded algebra $psu(2,2|4)$
\begin{equation}
T_AT_B-(-1)^{|A||B|}T_BT_A=
f_{AB}^{C}T_C \ . 
\end{equation}
The Killing form $\left<H_p.H_q\right>$
is also $\mathbf{Z}_4$ invariant so that
\begin{equation}
\left<\Omega(H_p),\Omega(H_q)\right>=
\left<H_p,H_q\right>
\end{equation}
that using (\ref{Omegaav})
leads to 
\begin{equation}\label{Hpq}
\left<H_p,H_q\right>=0 \ , \mathrm{unless}
 \  p+q=4 \ \mathrm{mod \ 4} \ . 
\end{equation}
We define $\left<\dots\right>$ in terms
of supertrace $\str$ where the supertrace
of the supermatrix $G$ (\ref{Gsup}) is defined as
\begin{equation}
\str(G)=\tr A-\tr B
\end{equation}
if $G$ is an even supermatrix and
\begin{equation}
\str(G)=\tr A+\tr B
\end{equation}
if $G$ is odd supermatrix. 
Note that the supertrace 
of two matrices $G,H$ obey
the relation
\begin{equation}
\str GH=(-1)^{|G||H|}\str HG \ , 
\end{equation}
where $|X|=1$ if $X$ is odd matrix.
This result will be important below
when we will discuss the variation
of the ghost action that
contains Grassmann odd matrices.

The  metric of the algebra is
defined as
\begin{equation}
G_{AB}=\str (T_AT_B) 
\end{equation}
and the  relation (\ref{Hpq}) implies
that the only nonzero elements are
$G_{[\uc\ud][\ue\uf]} \ , G_{\uc\ud} \ , 
G_{\alpha\hbeta}=-G_{\hbeta\alpha}$.
The structure constant of the 
$psu(2,2|4)$ algebra also possess the graded
anti-symmetry property
\begin{equation}
f_{AB}^DG_{DC}=
-(-1)^{|A||B|}f_{BA}^DG_{DC}=
-(-1)^{|B||C|}f_{AC}^DG_{DB} \ .
\end{equation}
The form of the action 
(\ref{Minaction})
is very useful since now
we can easily determine the
equation of motion from it.
Let us consider  the variation of
the group element $g$ as 
$\delta g=g\delta X$. 
Then the variation of the
current  
$J=g^{-1}dg$  is
equal to
\begin{equation}
\delta J_\mu=
-g^{-1}\delta g g^{-1}\partial_\mu g+
g^{-1}\partial_\mu\delta g
=\partial_\mu \delta X+[J_\mu,\delta X] \ .
\end{equation}
In order to find the variations 
of the currents $J^{(x)} \ ,
x=0,1,2,3$ we use the
relation (\ref{comhpq})
when  we also decompose
$\delta X$ as
$\delta X=\delta X^{(0)}+\delta X^{(1)}+
\delta X^{(2)}+\delta X^{(3)}$
where $\delta X^{(i)}\in
\mH_i$. Then we obtain
\begin{eqnarray}\label{rel}
\delta J^{(0)}=d \delta X^{(0)}+\com{J^{(0)},
\delta X^{(0)}}+
\com{J^{(1)},\delta X^{(3)}}+
\com{J^{(2)},\delta X^{(2)}}+
\com{J^{(3)},\delta X^{(1)}} \ , 
\nonumber \\
\delta J^{(1)}=
d\delta X^{(1)}+\com{J^{(0)},\delta
X^{(1)}}+
\com{J^{(1)},\delta X^{(0)}}+
\com{J^{(2)},\delta X^{(3)}}+
\com{J^{(3)},\delta X^{(2)}} \ ,
\nonumber \\
\delta J^{(2)}=
d\delta X^{(2)}+
\com{J^{(2)},\delta X^{(0)}}+
\com{J^{(1)},\delta X^{(1)}}
+\com{J^{(0)},\delta X^{(2)}}+
\com{J^{(3)},\delta X^{(3)}} \ ,
\nonumber \\
\delta J^{(3)}=
d\delta X^{(3)}+
\com{J^{(0)},\delta X^{(3)}}+
\com{J^{(1)},\delta X^{(2)}}
+\com{J^{(2)},\delta X^{(1)}}+
\com{J^{(3)},\delta X^{(0)}} \ .
\nonumber \\
\end{eqnarray}
Using these results we can
easily perform the variation of the
action (\ref{Minaction}) and  
we derive the equations
of motion 
\begin{eqnarray}\label{eqm}
\eta^{\mu\nu}\com{J^{(1)}_\mu,J^{(2)}_\nu}
+(\eta^{\mu\nu}+\frac{1}{2}\epsilon^{\mu\nu})
(\partial_\mu J^{(3)}_\nu
+[J^{(0)}_{\mu},J^{(3)}_\nu])
+(
\eta^{\mu\nu}+\frac{1}{2}\epsilon^{\mu\nu})
\com{J^{(2)}_\nu,J^{(1)}_\mu}+\nonumber \\
+\left(
[J_\nu^{(3)},N_\mu]P^{\mu\nu}+
[J_\nu^{(3)},\hN_\mu]\tilde{P}^{\mu\nu}
\right)=0 \ ,
\nonumber \\
\eta^{\mu\nu}
\com{J_\mu^{(3)},J^{(2)}_\nu}
+(\eta^{\mu\nu}+\frac{1}{2}\epsilon^{\mu\nu})
\com{J^{(3)}_\mu,J^{(2)}_\nu}+
(\eta^{\mu\nu}+\frac{1}{2}\epsilon^{\mu\nu})
(\partial_\nu J^{(1)}_\mu+
[J^{(0)}_\nu,J^{(1)}_\mu]+
\nonumber \\
+\left(
[J_\nu^{(1)},N_\mu]P^{\mu\nu}+
[J_\nu^{(1)},\hN_\mu]\tilde{P}^{\mu\nu}
\right)=0 \ , \nonumber \\
\partial_\mu J_\nu^{(2)}\eta^{\mu\nu}+
\eta^{\mu\nu}
[J^{(0)}_\mu,J^{(2)}_\nu]
+(\eta^{\mu\nu}+\frac{1}{2}\epsilon^{\mu\nu})
\com{J^{(3)}_\mu,J^{(3)}_\nu}
+(\eta^{\mu\nu}+\frac{1}{2}\epsilon^{\mu\nu})
\com{J^{(1)}_\nu,J^{(1)}_\mu}+
\nonumber \\
+\left(
[J_\nu^{(2)},N_\mu]P^{\mu\nu}+
[J_\nu^{(2)},\hN_\mu]\tilde{P}^{\mu\nu}
\right)=0 \ , \nonumber \\
\end{eqnarray}
where we have taken into
account the contribution
from the variation of the
ghost action
that is equal to
\begin{eqnarray}
\int d^2x \str(
\delta X^{(1)}
[\com{J^{(3)}_\nu,N_{\mu}}P^{\mu\nu}+
\com{J^{(3)}_\nu,\hN_{\mu}}\tP^{\mu\nu}]+
\nonumber \\
\delta X^{(2)}
[\com{J^{(2)}_\nu,N_{\mu}}P^{\mu\nu}+
\com{J^{(2)}_\nu,\hN_{\mu}}\tP^{\mu\nu}]+
\nonumber \\
\delta X^{(3)}
[\com{J^{(1)}_\nu,N_{\mu}}P^{\mu\nu}+
\com{J^{(1)}_\nu,\hN_{\mu}}\tP^{\mu\nu}]) \ . 
\nonumber \\
\end{eqnarray}
It is also important to
stress that  the ghost
variables transform non-trivially
for the gauge transformations
$\delta X^{(0)}\in \mH_0$  
\begin{eqnarray}
\delta \lambda=-\com{\delta X^{(0)},\lambda} \ ,
\delta w_{\mu}=-\com{\delta X^{(0)},w_{\mu}} \ ,
\nonumber \\
\delta \hlambda=
-\com{\delta X^{(0)},\hlambda} \ ,
\delta \hw_{\mu}=-\com{\delta X^{(0)},\hw_{\mu}} \ .
\nonumber \\
\end{eqnarray}
These relations also determine
the variation of the 
currents $N,\hN$ as
\begin{equation}
\delta N_\mu=\com{N_\mu,\delta X^{(0)}} \ ,
\delta \hN_{\mu}=\com{\hN_\mu,\delta X^{(0)}} \ .
\end{equation}
Using this fact it is easy
to see that the variation
of the action with respect
to $\delta X^{(0)}$ vanishes
and this result confirms the
 invariance of the action
under gauge transformations
from $SO(4,1)\times SO(5)$. 

We can  simplify the equations (\ref{eqm})
using the obvious identity
$dJ+J\wedge J=0$ that using
(\ref{comhpq}) can be
written as
\begin{eqnarray}\label{MCc}
\partial_\mu J^{(0)}_\nu-
\partial_\nu J^{(0)}_\mu +
\com{J^{(0)}_\mu,J^{(0)}_\nu}+
\com{J^{(1)}_\mu,J^{(3)}_\nu}+
\com{J^{(3)}_\mu,J^{(1)}_\nu}+
\com{J^{(2)}_\mu,J^{(2)}_\nu}=0 \ ,
\nonumber \\
\partial_\mu J^{(1)}_\nu-
\partial_\nu J^{(1)}_\mu +
\com{J^{(0)}_\mu,J^{(1)}_\nu}+
\com{J^{(1)}_\mu,J^{(0)}_\nu}+
\com{J^{(3)}_\mu,J^{(2)}_\nu}+
\com{J^{(2)}_\mu,J^{(3)}_\nu}=0 \ ,
\nonumber \\
\partial_\mu J^{(2)}_\nu-
\partial_\nu J^{(2)}_\mu +\com{J^{(1)}_\mu,
J^{(1)}_\nu}+\com{J^{(3)}_\mu,
J^{(3)}_\nu}+\com{J^{(0)}_\mu,J^{(2)}_\nu}+
\com{J^{(2)}_\mu,J^{(0)}_\nu}=0 \ ,
\nonumber \\
\partial_\mu J^{(3)}_\nu-
\partial_\nu J^{(3)}_\mu +
\com{J^{(0)}_\mu,J^{(3)}_\nu}+
\com{J^{(3)}_\mu,J^{(0)}_\nu}+
\com{J^{(1)}_\mu,J^{(2)}_\nu}+
\com{J^{(2)}_\mu,J^{(1)}_\nu}=0 \ .
\nonumber \\
\end{eqnarray}
For example, if we consider
the last equation in (\ref{MCc})
 in the form
 \begin{equation}
 \epsilon^{\mu\nu}
 D_{\mu}J^{(3)}_\nu=
 -\epsilon^{\mu\nu}
\com{J^{(1)}_\mu,J^{(2)}_\nu}
\end{equation}
and use it in  the  first equation
in  (\ref{eqm}) we obtain
\begin{equation}\label{eqm1}
\tP^{\mu\nu}
D_\mu J^{(3)}_\nu
+
[J_\nu^{(3)},N_\mu]P^{\mu\nu}+
[J_\nu^{(3)},\hN_\mu]\tilde{P}^{\mu\nu}
=0 \ .
\end{equation}
In the same we proceed with
the second equation in  (\ref{eqm})
and we get
\begin{equation}\label{eqm2}
P^{\mu\nu}D_\mu J^{(1)}_\nu
+
[J_\nu^{(1)},N_\mu]P^{\mu\nu}+
[J_\nu^{(1)},\hN_\mu]\tilde{P}^{\mu\nu}
=0 \ . 
\end{equation}
Finally, using (\ref{MCc})
we can show that  third equation in
(\ref{eqm}) implies 
\begin{equation}\label{eqm3a}
P^{\mu\nu}
D_\mu J^{(2)}_\nu
-\epsilon^{\mu\nu}
\com{J^{(1)}_\mu,J^{(1)}_\nu}
+[J_\nu^{(2)},N_\mu]P^{\mu\nu}+
[J_\nu^{(2)},\hN_\mu]\tilde{P}^{\mu\nu}
=0 \ 
\end{equation} 
and  
\begin{equation}\label{eqm3b}
\tP^{\mu\nu}
D_\mu J^{(2)}_\nu
+\epsilon^{\mu\nu}
\com{J^{(3)}_\mu,J^{(3)}_\nu}
+[J_\nu^{(2)},N_\mu]P^{\mu\nu}+
[J_\nu^{(2)},\hN_\mu]\tilde{P}^{\mu\nu}
=0 \ .
\end{equation}
In what follows we also need
the equation of motions for ghost
that take the form 
\begin{eqnarray}
P^{\mu\nu}\left(\partial_\nu \lambda
+\com{J^{(0)}_\nu,\lambda}\right)+
\frac{1}{2}
P^{\mu\nu}\com{\lambda,\hN_\nu}
+\frac{1}{2}\com{\lambda,\hN_\nu} \tP^{\nu\mu}=0 \ ,
\nonumber \\
\
\tilde{P}^{\mu\nu}\left(\partial_\nu \hlambda
+\com{J^{(0)}_\nu,\hlambda}\right)+
\frac{1}{2}
\com{\hlambda,N_\nu}P^{\nu\mu}
+\frac{1}{2}
\tP^{\mu\nu}\com{\hlambda,N_\nu}=0 \ . 
\nonumber \\
\end{eqnarray}

With the help of these
equations of motion we 
are ready to study the
properties of  BRST currents.
Firstly, let us consider
current 
\begin{equation}
j^\mu_R=\str(\hlambda P^{\mu\nu}
J^{(1)}_\nu) 
\end{equation}  
and calculate its divergence
\begin{eqnarray}
\partial_\mu j^\mu_R=
\str (\tP^{\nu\mu}\partial_\mu \hlambda
J^{(1)}_\nu+\hlambda P^{\mu\nu}
\partial_\mu J^{(1)}_\nu)=
\nonumber \\
\str(-\tP^{\nu\mu}[J^{(0)}_\mu,\hlambda]
-\tP^{\nu\mu}[\hlambda,N_\mu])
J^{(1)}_\nu+\nonumber \\
\str \hlambda(-P^{\mu\nu}
[J^{(0)}_\mu,J^{(1)}_\nu]-
\tP^{\mu\nu}[J^{(1)}_\mu,N_\nu]-
P^{\mu\nu}[J^{(1)}_\mu,\hN_\nu])=
\nonumber \\
=-P^{\mu\nu}\str \hlambda
\com{J^{(1)}_\mu,\hN_\nu}=
P^{\mu\nu}
\str J^{(1)}_\mu\com{\hlambda,\hN_\nu}
\nonumber \\
\end{eqnarray}
using $\tP^{\mu\nu}=P^{\nu\mu}$.
Now we can write 
this result as 
\begin{eqnarray}
\com{\hlambda,\hN_\mu}=
-\com{\hlambda\hlambda,\hw_\mu}=
-\frac{1}{2}\com{
\hlambda^{\halpha}\hlambda^{\hbeta}
(T_{\halpha} T_{\hbeta}+T_{\hbeta} T_{\halpha}),
\hw_\mu}
=\nonumber \\
-\frac{1}{2} \hlambda^{\halpha}
\lambda^{\hbeta} \gamma^{\uc}_{\halpha\hbeta}
\com{T_{\uc},\hw_\mu}=
-\frac{1}{2}\hlambda^{\halpha} (\gamma^{\uc})_{\halpha
\hbeta} \hlambda^{\hbeta} 
\com{T_{\uc},\hw_\mu} \ . 
 \nonumber \\
\end{eqnarray}
Now we immediately see that for
pure spinors this expression vanishes
and hence we obtain that the
current $j_{R}^\mu$ is conserved.

In the same way we can proceed
with the second current
\begin{equation}
j_{L}^\mu=\str(\lambda
\tP^{\mu\nu}J^{(3)}_\nu)
\end{equation}
and consider again its divergence
\begin{eqnarray}
\partial_\mu j^\mu_{L}&=&
\str\left(P^{\nu\mu}\partial_\mu \lambda 
J^{(3)}_\nu+
\lambda \tP^{\mu\nu}\partial_\mu J^{(3)}_\nu\right)=
\nonumber \\
&=&\str \left(\left
(-P^{\nu\mu}\com{J^{(0)}_\nu,\hlambda}-
P^{\nu\mu}\com{\hlambda,N_\mu}\right)
J^{(3)}_\nu +\right.
\nonumber \\
&+&\left.
\lambda\left(
-\tP^{\mu\nu}
\com{J^{(0)}_\mu, J^{(3)}_\nu}
-[J_\nu^{(3)},N_\mu]P^{\mu\nu}-
[J_\nu^{(3)},\hN_\mu]\tilde{P}^{\mu\nu}\right)
\right)=\nonumber \\
&=&\str P^{\mu\nu}
\com{\lambda,N_\mu}J^{(3)}_\nu \ . 
\nonumber \\
\end{eqnarray}
In the same way as above we
we can again show that this expression
is zero for pure spinor. Hence
we can define two conserved charges
\begin{eqnarray}\label{Q} 
Q_R=\int dx j^0_{R}=\int dx\str \hlambda P^{0\nu}
J^{(1)}_\nu
 \ , \nonumber \\
Q_L=\int dxj^0_{L}=\int dx 
\str\lambda \tP^{0\nu} J^{(3)}_\nu
\nonumber \\
\end{eqnarray}
that are time independent as follows
from the
analysis performed above. 
\subsection{Hamiltonian analysis}
In this section we present
the first step in the
Hamiltonian analysis of the
pure spinor action in $AdS_5\times S_5$.
It turns out that for these purposes
the  action
(\ref{Sha}) is very convenient. 
Firstly, using the explicit form
of the currents 
\begin{equation}
J^i_\mu=\partial_\mu Z^M E^i_M  \ , 
J^{\uc\ud}_\mu=
\partial_\mu Z^M\Omega^{[\uc\ud]}_M  
\end{equation}
we obtain
\begin{eqnarray}
\eta^{\mu\nu}
J^i_\mu G_{ij}J^j_\nu=
\eta^{\mu\nu}
\partial_\mu Z^ME_M^{i}G_{ij}\partial_\nu Z^N
E_N^{j} \ ,  
\nonumber \\
\epsilon^{\mu\nu}J^i_\mu J^jB_{ij}=
\epsilon^{\mu\nu}
\partial_\mu Z^M\partial_\nu Z^N
B_{MN}\ , \nonumber \\
\end{eqnarray}
where 
\begin{eqnarray}
B_{MN}=(-1)^{|N|M+i|}E_M^iB_{ij}E^j_N=
\nonumber \\
=-(-1)^{|M||N|}B_{NM} \ .
\nonumber \\
\end{eqnarray}
Then the action (\ref{Sha}) takes the form
\begin{eqnarray}
S=-\int d^2x\sqrt{-\eta}
(\frac{1}{2}\eta^{\mu\nu}\partial_\mu 
Z^ME_M^iG_{ij}\partial_\nu Z^N
E_N^j
-\frac{\epsilon^{\mu\nu}}{2}
\partial_\mu Z^M\partial_\nu Z^N
B_{MN}
\nonumber \\
+w_{\mu\alpha}P^{\mu\nu}
\partial_\nu \lambda^\alpha
+\hw_{\mu\halpha}\tP^{\mu\nu}
\partial_\nu\hlambda^{\halpha}
+\nonumber \\
+N_{\uc\ud\mu}
P^{\mu\nu}\partial_MZ^M
\Omega^{[\uc\ud]}_M
+\hN_{\uc\ud\mu}
\tP^{\mu\nu}\partial_MZ^M
\Omega^{[\uc\ud]}_M-
N_{cd\mu}P^{\mu\nu}\hN^{cd}_\nu+N_{c'd'\mu}P^{\mu\nu}
N^{c'd'}_\nu) \ .   
\nonumber \\
\end{eqnarray}
Then we easily obtain the
momenta $P_M,\pi_{\alpha} \ , 
\hpi_{\halpha}$ as
\begin{eqnarray}\label{PM}
P_M=\frac{\delta^L S}{\delta^L
\partial_0 Z^M}=
E_M^iG_{ij}\partial_0Z^NE_N^j
+\partial_1 Z^NB_{MN}
-\nonumber \\
-N_{\uc\ud\mu}P^{\mu 0}
\Omega^{[\uc\ud]}_M-\hN_{\uc\ud\mu}
\tP^{\mu 0}\Omega^{[\uc\ud]}_M \ , 
\nonumber \\ 
\pi_{\alpha}=\frac{\delta S}{\delta 
\partial_0 \lambda^\alpha}=
-w_{\alpha\mu}P^{\mu 0} \ , 
\hpi_{\halpha}=\frac{\delta S}{
\delta \partial_0\hlambda^{\halpha}}=
-\hw_{\halpha\mu}\tP^{\mu 0} \ .
\nonumber \\ 
\end{eqnarray}
Our goal  is to express $J_0^i$ using
the canonical variables $P_M , \ Z^N$
and  $\lambda,\hlambda,\pi,\hpi$.
 In fact, it is easy to
see that (\ref{PM}) implies
\begin{eqnarray}\label{JPM}
J^i_0=
G^{ij}E_j^M(P_M-\partial_1 Z^NB_{NM}
-\nonumber \\
-\frac{1}{2}
\pi_{\alpha}(\gamma_{\uc\ud})^\alpha
_\beta\lambda^\beta
\Omega^{[\uc\ud]}_M-\frac{1}{2}
\hpi_{\halpha}(\gamma_{\uc\ud})^{\halpha}_{\hbeta}
\hlambda^{\hbeta}\Omega^{[\uc\ud]}_M)
\equiv G^{ij}E_j^M\bP_M
\ . 
\nonumber \\  
\end{eqnarray}
Now we are ready to 
study the action 
BRST $Q_{R,L}$ (\ref{Q})
on the  
 ghost number $(1,1)$ function
 $F$
\begin{equation}\label{F}
F(Z,\lambda,\hlambda)=
\lambda^\alpha\hlambda^{\halpha}
V_{\alpha\halpha}(Z)
\end{equation}
that depends on $Z$ only
 and as we will show 
 it is related to the massless
state of the string in $AdS_5\times
S_5$ background. 
As the first step we use
(\ref{JPM}) to express
the operator (\ref{Q}) 
as 
\begin{eqnarray}
Q_R=\int dx
\hlambda^{\halpha}
G_{\halpha\beta} P^{0\mu}
J^{\beta}_\mu=
-\int dx \hlambda^{\halpha}
G_{\halpha\beta}[G^{\beta\halpha}
E^M_{\halpha}
\hP_M+
J^{\beta}_1] \ ,
\nonumber \\
Q_L=\int dx \lambda^\alpha G_{\alpha\hbeta}
\tP^{0\mu}J_\mu^{\hbeta}=
-\int dx^1\lambda^\alpha
G_{\alpha\hbeta}[G^{\hbeta\beta}
E_\beta^M\hP_M-J^{\hbeta}_1] \ . 
\nonumber \\
\end{eqnarray}
Since the function $F$
given in (\ref{F})
does not depend on $P_M$
the Poisson bracket
of $F$ with 
$J^\alpha_1 \ , J^{\halpha}_1$
vanishes. 
 Then
using the Poisson brackets
\begin{eqnarray}
\pb{\bP_M(x),Z^N(y)}=-
(-1)^{|M|}\delta^N_M\delta(x-y) \ , 
\nonumber \\ 
\pb{\bP_M(x),\lambda^\alpha(y)}=
\frac{1}{2}(\gamma_{\uc\ud})
^\alpha_\beta\lambda^\beta 
\Omega^{[\uc\ud]}_M\delta(x-y) \ ,
\nonumber \\
\pb{\bP_M(x),\hlambda^{\halpha}(y)}=
\frac{1}{2}(\gamma_{\uc\ud})
^{\halpha}_{\hbeta}\hlambda^{\hbeta} 
\Omega^{[\uc\ud]}_M\delta(x-y) \ 
\nonumber \\
\end{eqnarray}
we can easy  calculate
the Poisson bracket  between  
$Q_{R,L}$ and $F$
\begin{eqnarray}
\pb{Q_L,F(y)}=
\lambda^\gamma
E_\gamma^M
[\lambda^\alpha\lambda^{\halpha}
(-1)^{|M|}\partial_M V_{\alpha\halpha}-
\nonumber \\
-\frac{1}{2}
\lambda^{\halpha}(\gamma_{\uc\ud})^\alpha
_\beta\lambda^\beta\Omega^{[\uc\ud]}_M V_{\alpha\halpha}
-\frac{1}{2}(\gamma_{\uc\ud})
^{\halpha}_{\hbeta}\hlambda^{\hbeta} 
\Omega^{[\uc\ud]}_MV_{\alpha\halpha}]=
\nonumber \\
\lambda^\gamma\lambda^\alpha
\lambda^{\halpha}
E_\gamma^M[
(-1)^{|M|}\partial_M V_{\alpha\halpha}-
\frac{1}{2}(\gamma_{\uc\ud})^\beta_\alpha
\Omega^{[\uc\ud]}_MV_{\beta\halpha}-
\frac{1}{2}(\gamma_{\uc\ud})^{\hbeta}_{\halpha}
\Omega_M^{[\uc\ud]}V_{\alpha\hbeta}]=
\nonumber \\
=\lambda^\gamma\lambda^\alpha
\lambda^{\halpha}\nabla_{\gamma}V_{\alpha\halpha} \ .
\end{eqnarray}
In the same way we can
calculate the action of $Q_R$ on $V$ and
we get the result
\begin{eqnarray}
\pb{Q_R,F(y)}=
\hlambda^{\hgamma}\lambda^{\alpha}
\hlambda^{\halpha}
E_{\hgamma}^M[
(-1)^{|M|}\partial_M V_{\alpha\halpha}-
\frac{1}{2}(\gamma_{\uc\ud})^\beta_\alpha
\Omega^{[\uc\ud]}_MV_{\beta\halpha}-
\frac{1}{2}(\gamma_{\uc\ud})^{\hbeta}_{\halpha}
\Omega_M^{[\uc\ud]}V_{\alpha\hbeta}]=
\nonumber \\
=\hlambda^{\hgamma}\lambda^{\alpha}
\hlambda^{\halpha}\nabla_{\hgamma}V_{\alpha\halpha}
\nonumber \\
\end{eqnarray}
Now using the 
relations (\ref{purespinorcon})
we obtain from the requirements
that $F$ is in the cohomology of $Q_R,Q_L$
\begin{equation}
\pb{Q_L,F}=\pb{Q_R,F}=0
\end{equation}
the equations  
\begin{equation}\label{gammam}
\gamma^{\gamma\alpha}_{mnpqr}
\nabla_{\gamma}V_{\alpha\halpha}=
\gamma^{\hgamma\halpha}_{mnpqr}
\nabla_{\hgamma}V_{\alpha\halpha}=0 \ . 
\end{equation}
As was shown in \cite{Berkovits:2004xu}
 these equations 
correctly describe the on-shell
fluctuations around the $AdS_5\times S_5$
background. It is one of the main
results of this paper that we were
able to determine the action 
of the BRST operators $Q_{L,R}$ on
the function $F$ from the first principles
of the classical canonical formalism. 

Generally, if we consider function
of the ghost number $(n,m)$ in the form
\begin{equation}
H^{(n,m)}(\lambda,\hlambda,Z)=
\lambda^{\alpha_1}\dots\lambda^{\alpha_m}
\hlambda^{\hbeta_1}\dots\hlambda^{\halpha_n}
A_{\alpha_1\dots\alpha_m\hbeta_1\dots\hbeta_n}(Z) \ 
\end{equation}
we easily obtain that the action of $Q_L,Q_R$
on it takes the form
\begin{eqnarray}\label{generaq}
\pb{Q_L,H^{(n,m)}}=
\lambda^{\kappa}\lambda^{\alpha_1}\dots\lambda^{\alpha_m}
\hlambda^{\hbeta_1}\dots\hlambda^{\halpha_n}
\nabla_{\kappa}A_{\alpha_1\dots\alpha_m\hbeta_1\dots\hbeta_n} \ , 
\nonumber \\
\pb{Q_R,H^{(n,m)}}=
\hlambda^{\hkappa}
\lambda^{\alpha_1}\dots\lambda^{\alpha_m}
\hlambda^{\hbeta_1}\dots\hlambda^{\halpha_n}
\nabla_{\hkappa}A_{\alpha_1\dots\alpha_m\hbeta_1\dots\hbeta_n} \ ,
\nonumber \\ \ 
\end{eqnarray}
where 
\begin{eqnarray}
\nabla_{\kappa}A_{\alpha_1\dots\alpha_m\hbeta_1\dots\hbeta_n}=
E_\kappa^M((-1)^{|M|}
\partial_MA_{\alpha_1\dots\alpha_m\hbeta_1\dots\hbeta_n}
-\Omega^{[\uc\ud]}\nabla_{[\uc\ud]}
A_{\alpha_1\dots\alpha_m\hbeta_1\dots\hbeta_n}) \ , 
\nonumber \\
\nabla_{\hkappa}A_{\alpha_1\dots\alpha_m\hbeta_1\dots\hbeta_n}=
E_{\hkappa}^M((-1)^{|M|}
\partial_MA_{\alpha_1\dots\alpha_m\hbeta_1\dots\hbeta_n}
-\Omega^{[\uc\ud]}\nabla_{[\uc\ud]}
A_{\alpha_1\dots\alpha_m\hbeta_1\dots\hbeta_n}) \ , 
\nonumber \\
\nabla_{[\uc\ud]}
A_{\alpha_1\dots\alpha_m\hbeta_1\dots\hbeta_n}=\frac{1}{2}
(\gamma_{cd})_{\alpha_1}^{\gamma}
A_{\gamma\alpha_2\dots\alpha_m\hbeta_1\dots\hbeta_n}+\dots 
+(\gamma_{cd})_{\alpha_m}^{\gamma}
A_{\alpha_1\dots\alpha_{m-1}\gamma\hbeta_1\dots\hbeta_n}
+\nonumber \\
\frac{1}{2}
(\gamma_{cd})_{\hbeta_1}^{\hgamma}
A_{\alpha_1\dots\alpha_m\hgamma \hbeta_2\dots\hbeta_n}+\dots 
+\frac{1}{2}
(\gamma_{cd})_{\hbeta_n}^{\hgamma}
A_{\alpha_1\dots\alpha_m\hbeta_1\dots\hbeta_{n-1}\hgamma} \ . 
\nonumber \\
\end{eqnarray}
Now we will study the 
consequence of  the gauge invariance
of the function $F$ 
\begin{eqnarray}\label{deltaF}
\delta F=
\pb{Q_L,\Lambda}+\pb{Q_R,\hat{\Lambda}} \ ,  \nonumber \\
\pb{Q_R,\Lambda}=\pb{Q_L,\hat{\Lambda}}=0 \ , 
\nonumber \\
\end{eqnarray}
where 
\begin{equation}
\Lambda(\hlambda,Z)=\hlambda^{\halpha}\Gamma_{\halpha}(Z) \ , 
\hat{\Lambda}(\lambda,Z)=\lambda^{\alpha}\hat{\Gamma}_{\alpha}(Z) \ .
\end{equation}
Using (\ref{generaq}) we easily get
\begin{equation}\label{pbqL}
\pb{Q_L,\Lambda}=\lambda^\alpha\hlambda^{\hbeta}
\nabla_\alpha \Gamma_{\hbeta} \ , 
\pb{Q_R,\hat{\Lambda}}=\lambda^{\alpha}\hlambda^{\hbeta}
\nabla_{\hbeta}\Gamma_{\alpha} \ 
\end{equation}
and if we define $\delta F=
\lambda^\alpha \hlambda^{\hbeta}
\delta V_{\alpha\hbeta}$ we obtain
from (\ref{deltaF}) and (\ref{pbqL})
following transformations rule of $V_{\alpha\hbeta}$
\begin{equation}
\delta V_{\alpha\hbeta}=
\nabla_{\alpha}\Gamma_{\hbeta}+
\nabla_{\hbeta}\hat{\Gamma}_{\alpha} \ .
\end{equation}
Finally, using 
(\ref{generaq}) and (\ref{purespinorcon})
 the expressions on the second line
in (\ref{deltaF}) 
imply
\begin{equation}
\gamma^{\alpha\beta}_{mnpqr}
\nabla_{\alpha}\Gamma_\beta=0 \ , 
\gamma^{\halpha\hbeta}_{mnpqr}
\nabla_{\halpha}\hat{\Gamma}_{\hbeta}=0 \ .
\end{equation}
In summary, we have derived
using the canonical Hamiltonian
formalism  that the
the ghost number $(1,1)$ 
function $F(\lambda,\hlambda,Z)$
defined on extended phase
space  
 describes the massless
fluctuations on the $AdS_5\times
S_5$ background. 
We hope that this result 
can serve as an additional support for
 the analysis performed in 
paper in \cite{Berkovits:2004xu}.
\\
\\
{\bf Acknowledgement}
\\
\\
I would like to thank M. Bianchi for
many stimulating discussions and for his support
of my work. I would also like to thank N. Berkovits
and P. A.  Grassi for very interesting correspondence.
This work
 was supported in part by the Czech Ministry of
Education under Contract No. MSM
0021622409, by INFN, by the MIUR-COFIN
contract 2003-023852, by the EU
contracts MRTN-CT-2004-503369 and
MRTN-CT-2004-512194, by the INTAS
contract 03-516346 and by the NATO
grant PST.CLG.978785.

\newpage

\end{document}